\documentstyle[11pt,paspconf,epsf]{article}

\def\vecp{\mbox{\boldmath $p$}}
\def\vecq{\mbox{\boldmath $q$}}
\def\vecr{\mbox{\boldmath $r$}}
\def\vecv{\mbox{\boldmath $v$}}
\def\vecw{\mbox{\boldmath $w$}}
\def\vecF{\mbox{\boldmath $F$}}
\def\vecFt{\mbox{\boldmath $\tilde F$}}

\begin{document}

\title{Variable Timestep Integrators for Long-Term Orbital
       Integrations\footnotemark}
\footnotetext{To appear in Computational Astrophysics, Proc.\ 12th Kingston
              Meeting, ed.\ D.~A.~Clarke \& M.~J.~West (San Francisco:
              Astronomical Society of the Pacific).}

\author{Man Hoi Lee and Martin J.~Duncan}
\affil{Department of Physics, Queen's University, Kingston,
       ON K7L 3N6, Canada}

\author{Harold F.~Levison}
\affil{Space Science Department, Southwest Research Institute, Boulder,
       CO~80302, USA}

\begin{abstract}
Symplectic integration algorithms have become popular in recent years in
long-term orbital integrations because these algorithms enforce certain
conservation laws that are intrinsic to Hamiltonian systems.
For problems with large variations in timescale,
it is desirable to use a variable timestep.
However, naively varying the timestep destroys the desirable properties of
symplectic integrators.
We discuss briefly the idea that choosing the timestep in a time symmetric
manner can improve the performance of variable timestep integrators.
Then we present a symplectic integrator which is based on decomposing the
force into components and applying the component forces with different
timesteps.
This multiple timescale symplectic integrator has all the desirable
properties of the constant timestep symplectic integrators.
\end{abstract}


\section{Symplectic Integrators}

Long-term numerical integrations play an important role in our understanding
of the dynamical evolution of many astrophysical systems (see, e.g., Duncan,
these proceedings, for a review of solar-system integrations).
An essential tool for long-term integrations is a fast and accurate
integration algorithm.
Symplectic integration algorithms (SIAs) have become popular in recent years
because the Newtonian gravitational $N$-body problem is a Hamiltonian problem
and SIAs enforce certain conservation laws that are intrinsic to Hamiltonian
systems (see Sanz-Serna \& Calvo 1994 for a general introduction to SIAs).

For an autonomous Hamiltonian system, the equations of motion are
\begin{equation}
{d\vecw / dt} = \{\vecw,H\},
\label{Heq}
\end{equation}
where $H(\vecw)$ is the explicitly time-independent Hamiltonian,
$\vecw = (\vecq,\vecp)$ are the $2d$ canonical phase-space coordinates,
$\{\ ,\ \}$ is the Poisson bracket, and $d$($= 3N$) is the number of degrees
of freedom.
The formal solution of Eq.(\ref{Heq}) is
\begin{equation}
\vecw(t) = \exp({t \{\ ,H\}}) \vecw(0).
\end{equation}
If the Hamiltonian $H$ has the form $H_A + H_B$, where $H_A$ and $H_B$ are
separately integrable, we can devise a SIA of constant timestep $\tau$
by approximating $\exp(\tau \{\ ,H\})$ as a composition of terms like
$\exp(\tau \{\ ,H_A\})$ and $\exp(\tau \{\ ,H_B\})$.
For example, a second-order SIA is
\begin{equation}
\exp({\tau \over 2} \{\ ,H_A\}) \exp(\tau \{\ ,H_B\})
\exp({\tau \over 2} \{\ ,H_A\}) .
\label{SIA2}
\end{equation}

For the gravitational $N$-body problem, we can write
$H = T(\vecp) + V(\vecq)$, where $T(\vecp)$ and $V(\vecq)$ are the kinetic
and potential energies.
Then the second-order SIA Eq.(\ref{SIA2}) becomes
\begin{equation}
\vecp_{n+{1\over2}} = \vecp_n + {\tau \over 2} \vecF(\vecq_n) ,
\quad
\vecq_{n+1} = \vecq_n + \tau \vecv(\vecp_{n+{1\over2}}) ,
\quad
\vecp_{n+1} = \vecp_{n+{1\over2}} + {\tau \over 2} \vecF(\vecq_{n+1}) ,
\label{LF}
\end{equation}
where $\vecF = -\partial V/\partial \vecq$ and $\vecv =
\partial T/\partial \vecp$; this is the familiar leapfrog integrator.
For solar-system type integrations, a central body (the Sun) is much more
massive than the other bodies in the system, and it is better to write
$H = H_{\rm Kep} + H_{\rm int}$, where $H_{\rm Kep}$ is the part of the
Hamiltonian that describes the Keplerian motion of the bodies around the
central body and $H_{\rm int}$ is the part that describes the perturbation
of the bodies on one another.
Symplectic integrators using this decomposition of the Hamiltonian were
introduced by Wisdom \& Holman (1991),
and they are commonly called mixed variable symplectic (MVS) integrators.

The constant timestep SIAs have the following desirable properties:

\noindent
(1) As their names imply, SIAs are symplectic, i.e., they preserve
$d\vecp \wedge d\vecq$.

\noindent
(2) For sufficiently small $\tau$, SIAs solve almost exactly a nearby
``surrogate'' autonomous Hamiltonian problem with
${\tilde H} = H + H_{\rm err}$.
For example, the second-order SIA in Eq.(\ref{SIA2}) has
\begin{equation}
H_{\rm err} = {\tau^2 \over 12} \{\{H_A,H_B\}, H_B + {1 \over 2} H_A\}
              + O(\tau^4) .
\label{Herr}
\end{equation}
Consequently, we expect that the energy error is bounded and the position
(or phase) error grows linearly.

\noindent
(3) Many SIAs (e.g., Eq.[\ref{SIA2}]) are time reversible.
Note, however, that there are algorithms that are symplectic but not
reversible.

Fig.~1 shows the energy error $\Delta E/E$ of an integration of the
$e = 0.5$ Kepler orbit using the constant timestep leapfrog integrator
Eq.(\ref{LF}).
(In this and all subsequent Figures, the orbits are initially at the
apocenter $r_a = a (1 + e)$,
where $a$ and $e$ are the semi-major axis and eccentricity.)
Fig.~1 illustrates that there is no secular drift in $\Delta E/E$.

\begin{figure}
\centering \leavevmode \epsfxsize=.45\textwidth \epsfbox{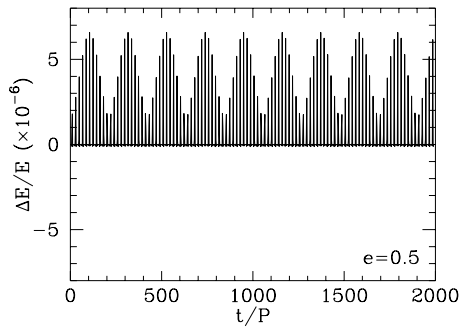}
\caption{Energy error $\Delta E/E$ of an integration of the Kepler
problem using the constant timestep leapfrog integrator Eq.(\ref{LF}).
The eccentricity $e = 0.5$, the timestep $\tau = P/4000$, and $P$ is the
orbital period.}
\end{figure}

\section{Simple Variable Timestep}

For problems with large variations in timescale (due to close encounters or
high eccentricities), it is desirable to use a variable timestep.
A common practice is to set the timestep using the phase-space coordinates
$\vecw_n$ at the beginning of the timestep: $\tau = h(\vecw_n)$.
If the SIAs discussed in \S~1 are implemented with this simple variable
timestep scheme,
they are still symplectic if we assume that the sequence of timesteps
determined for a particular initial condition are also used to integrate
neighboring initial conditions (see Skeel \& Gear 1992 for another point
of view).
However, tests have shown that this and similar simple variable timestep
schemes [and also $\tau = h(t)$] destroy the desirable properties of the
integrators (e.g., Gladman, Duncan, \& Candy 1991; Calvo \& Sanz-Serna 1993).
In Fig.~2{\it a} we show the energy error of an integration with $e = 0.5$
and $h(r) = \tau_0 (r/r_a)^{3/2}$.
Although the error is initially smaller than that in Fig.~1 (the integrations
shown in Figs.~1 and 2 use nearly the same number of timesteps per orbit),
it shows a linear drift.
In Fig.2{\it a} we also show the errors for two neighboring initial
conditions ($e = 0.5 \pm 0.005$).
They were integrated using the timesteps determined for the $e = 0.5$ orbit.
Note that these errors grow even faster.

\begin{figure}
\plottwo{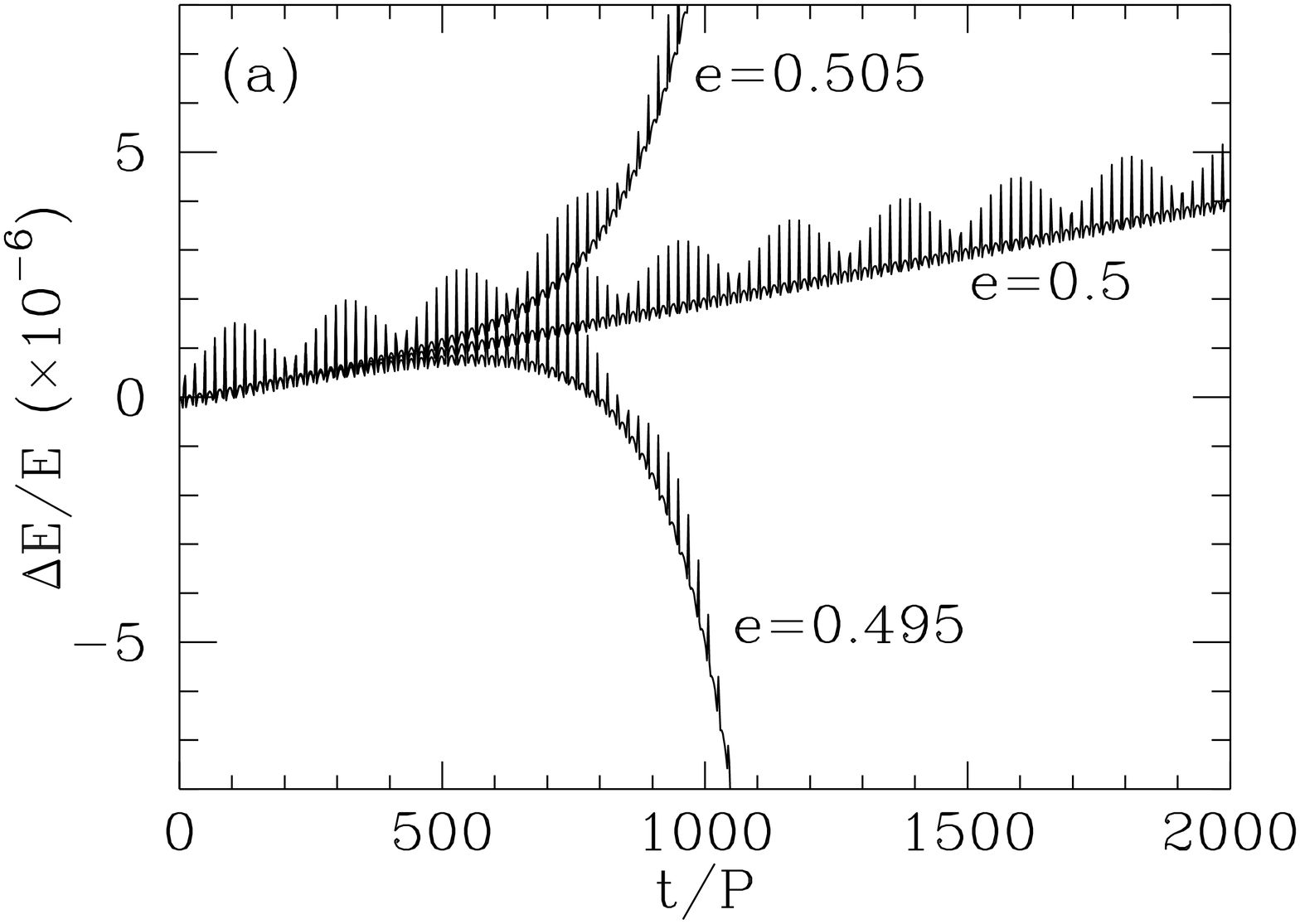}{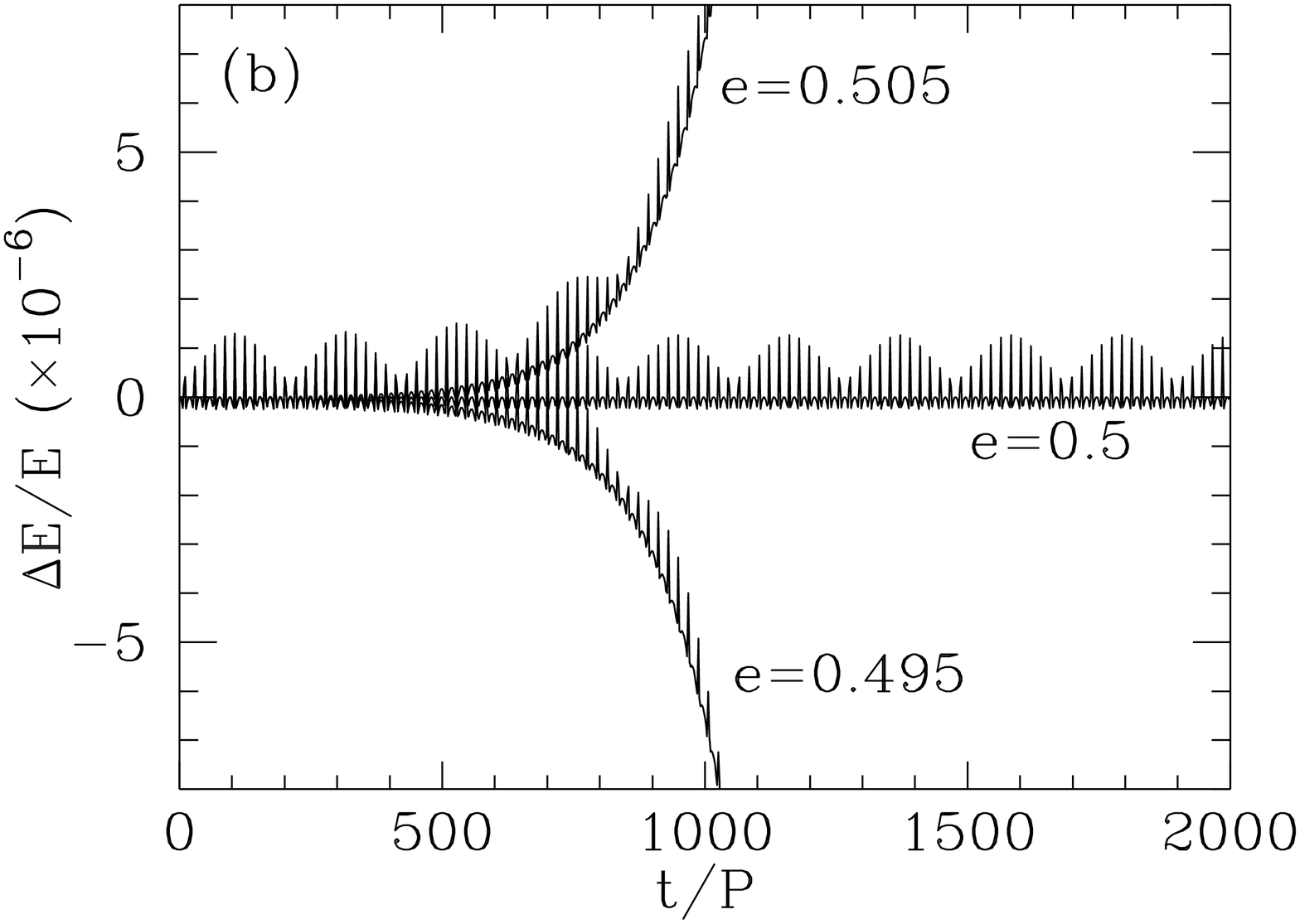}
\caption{({\it a}) Same as Fig.~1, but using the leapfrog integrator
Eq.(\ref{LF}) with the simple variable timestep $\tau = \tau_0 (r/r_a)^{3/2}$,
where $\tau_0 = P/2000$.
The errors for two neighboring initial conditions are also shown (see text).
({\it b}) Same as ({\it a}), but using the symmetrized timestep criterion.}
\end{figure}

The degradation in performance is due to the fact that the properties (2) and
(3) listed in \S~1 are no longer true.
The algorithm is not time reversible because the timestep depends only on the
coordinates at the beginning of the timestep.
Since $H_{\rm err}$ depends explicitly on the timestep $\tau$ (see, e.g.,
Eq.[\ref{Herr}]),
a variable timestep changes the surrogate Hamiltonian problem that the
integrator is solving from step to step.
Thus the solution from $t=0$ to $t_n$ after $n$ steps is not in general
the solution of a nearby autonomous Hamiltonian problem.

\section{Time Symmetrization}

Hut, Makino, \& McMillan (1995; see also Funato et al.~1996; Hut, these
proceedings) pointed out that the performance of a variable timestep
integrator can be improved if time reversibility is restored by choosing the
timestep in a time symmetric manner: $\tau = \tau(\vecw_n,\vecw_{n+1})$
with $\tau(\vecw_n,\vecw_{n+1}) = \tau(\vecw_{n+1},\vecw_n)$.
For example, $\tau = [h(\vecw_n) + h(\vecw_{n+1})]/2$.
We have tested this idea in detail by (i) integrating a series of
problems (the pendulum, Kepler orbits, and the restricted $3$-body
problem) using a symmetrized variable timestep leapfrog integrator and
(ii) integrating the restricted $3$-body problem using a symmetrized MVS
integrator.
We have found that time symmetrization usually reduces the drift in energy
(or the Jacobi constant) to negligible levels.
Fig.~2{\it b} shows an integration of the $e = 0.5$ Kepler orbit using
the symmetrized leapfrog integrator,
and it should be compared to Fig.~2{\it a}.

In Fig.~2{\it b} we also show the errors for two neighboring initial
conditions that were integrated using the timesteps determined for the
$e = 0.5$ orbit.
There is no improvement in the error growth for the neighboring initial
conditions.
Integrating an orbit using the timesteps determined for another orbit
may seem to be somewhat artificial,
but similar situations do occur in realistic integrations.
For example, if we integrate an $N$-body system using a shared timestep
scheme, the timestep used by all of the particles is determined by the
few having the strongest close encounter.
In practice, since the symmetrized timestep criterion depends on $\vecw_{n+1}$,
a symmetrized integrator also has the disadvantage that it requires iteration
and can be significantly slower than the original integrator
(unless the original integrator is also implicit).

\section{Multiple Timescale Symplectic Integrators}

In this section we describe a ``variable timestep'' SIA that has all the
desirable properties of the constant timestep SIAs.
The algorithm is based on ideas proposed by Skeel \& Biesiadecki (1994;
see also MacEvoy \& Scovel 1994).
It is also similar to the individual timestep scheme of Saha \& Tremaine
(1994).

For simplicity, let us consider in particular the Kepler problem with
$T(\vecp) = |\vecp|^2/2$ and $V(\vecr) = -1/r$.
We choose a set of cutoff radii $r_1 > r_2 > \cdots$ and decompose the
potential $V$ into $V_i$, or equivalently the force $\vecF$ into
$\vecF_i = -\partial V_i/\partial \vecr$, such that
(i) $\vecF = \sum_{i=0}^\infty \vecF_i$,
(ii) $\vecF_i$ (except $\vecF_0$) is zero at $r > r_i$, and
(iii) $\vecF_i$ is ``softer'' than $\vecF_{i+1}$.
The force $\vecF_i$ is to be applied with a timestep $\tau_i$.
If we assume that $\tau_i/\tau_{i+1} = M_{i+1}$ is an integer,
we can apply the second-order SIA in Eq.(\ref{SIA2}) recursively to obtain
the following second-order algorithm:
\begin{eqnarray}
\exp(\tau_0 \{\ ,H\})\!\!
&\approx&\!\! \exp({\tau_0 \over 2} K_0)\,
              \exp[\tau_0 (D + K_1 + K_2 + \cdots)]\,
              \exp({\tau_0 \over 2} K_0) \nonumber \\
&\approx&\!\! \exp({\tau_0 \over 2} K_0)\, \biggl[
              \exp({\tau_1 \over 2} K_1)\,
              \exp[\tau_1 (D + K_2 + K_3 + \cdots)] \nonumber \\
&&\quad \times \exp({\tau_1 \over 2} K_1)\biggr]^{M_1}\,
               \exp({\tau_0 \over 2} K_0) \label{VSIA2} \\
&\vdots& \nonumber
\end{eqnarray}
where $D = \{\ ,T\}$ and $K_i = \{\ ,V_i\}$.
Hereafter, we adopt $M_i = 2$.

The multiple timescale algorithm Eq.(\ref{VSIA2}) has an overall timestep
$\tau_0$, but it is effectively a variable timestep scheme because the
recursion terminates at level $i$ if $K_{i+1} + K_{i+2} + \cdots = 0$ during
a substep of length $\tau_i$.
For example, if the particle is in the region $r_1 > r > r_2$ during an
overall step, Eq.(\ref{VSIA2}) reduces to
\begin{equation}
\exp({\tau_0 \over 2} K_0 + {\tau_0 \over 4} K_1)\, \exp({\tau_0 \over 2} D)\,
\exp({\tau_0 \over 2} K_1)\, \exp({\tau_0 \over 2} D)\,
\exp({\tau_0 \over 4} K_1 + {\tau_0 \over 2} K_0) .
\end{equation}
Since Eq.(\ref{VSIA2}) is based on the recursive application of
Eq.(\ref{SIA2}),
it has all the desirable properties of a constant timestep SIA.
It is obviously symplectic and time reversible.
We can also derive the surrogate autonomous Hamiltonian solved by this
integrator.
For example, if we decompose $V$ into two levels $V_0$ and $V_1$ only,
the error Hamiltonian is (cf.~Eq.[\ref{Herr}])
\begin{eqnarray}
H_{\rm err}\!\! &=&\!\!
{\tau_0^2 \over 12} \{\{V_0,T\}, T + {1 \over 2} V_0\} +
{\tau_1^2 \over 12} \{\{V_1,T\}, T + {1 \over 2} V_1\} +
{\tau_0^2 \over 12} \{\{V_0,T\}, V_1\} \nonumber \\
&& \quad {} + O(\tau_0^4) .
\end{eqnarray}

\begin{figure}
\centering \leavevmode \epsfxsize=.45\textwidth \epsfbox{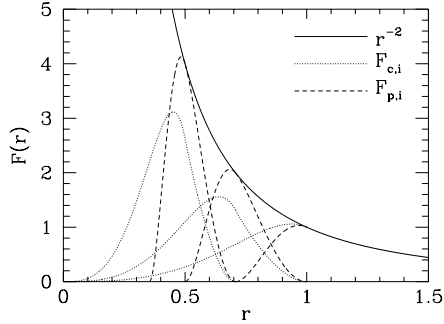}
\caption{Decomposition of the force $F(r) = 1/r^2$ into $F_{c,i}$ or
$F_{p,i}$.
The $i = 0$, $1$, and $2$ components, with $r_1 = 1$ and
$r_i/r_{i+1} = \protect\sqrt{2}$, are shown.}
\end{figure}

After some experiments, we found two force decompositions that work well.
If we write $\vecF_{c,0} = \vecFt_{c,0}$ and $\vecF_{c,i} = \vecFt_{c,i} -
\vecFt_{c,i-1}$ for $i \neq 0$, one of the decompositions is
\begin{equation}
\vecFt_{c,i-1} = \left\{\begin{array}{ll}
-\vecr/r^3 & \mbox{if $r \ge r_i$,} \\
- \left[9 \left(r/r_i\right)^2 - 5 \left(r/r_i\right)^6\right] \vecr/4 r_i^3
& \mbox{if $r < r_i$.}
\end{array}\right.
\end{equation}
An alternative decomposition uses
\begin{equation}
\vecFt_{p,i-1} = \left\{\begin{array}{ll}
-\vecr/r^3 & \mbox{if $r \ge r_i$,} \\
- f\!\left(r_i - r \over r_i - r_{i+1}\right) \vecr/r^3
& \mbox{if $r_{i+1} \le r < r_i$,} \\
0 & \mbox{if $r < r_{i+1}$,}
\end{array}\right.
\end{equation}
where $f(x) = 2 x^3 - 3 x^2 +1$.
Unlike the forces suggested by Skeel \& Biesiadecki (1994),
these forces have continuous first derivatives and decrease rapidly (or
exactly) to zero at $r \ll r_i$ (see Fig.~3).
We shall not provide the details here,
but we can understand why these properties are important from an analysis
of $H_{\rm err}$.
In Fig.~4 we show the energy error $\Delta E/E$ of two integrations
using the multiple timescale symplectic integrator with $\vecF_{c,i}$.
As expected, there is no secular drift in $\Delta E/E$.
Note also that, with the chosen integration parameters
($\tau_i \propto r_i^2$), the maximum error is almost independent of the
pericentric distance (which changes by $10^3$ in the two cases shown).
This again agrees with the expectation from an analysis of $H_{\rm err}$.

\begin{figure}
\plottwo{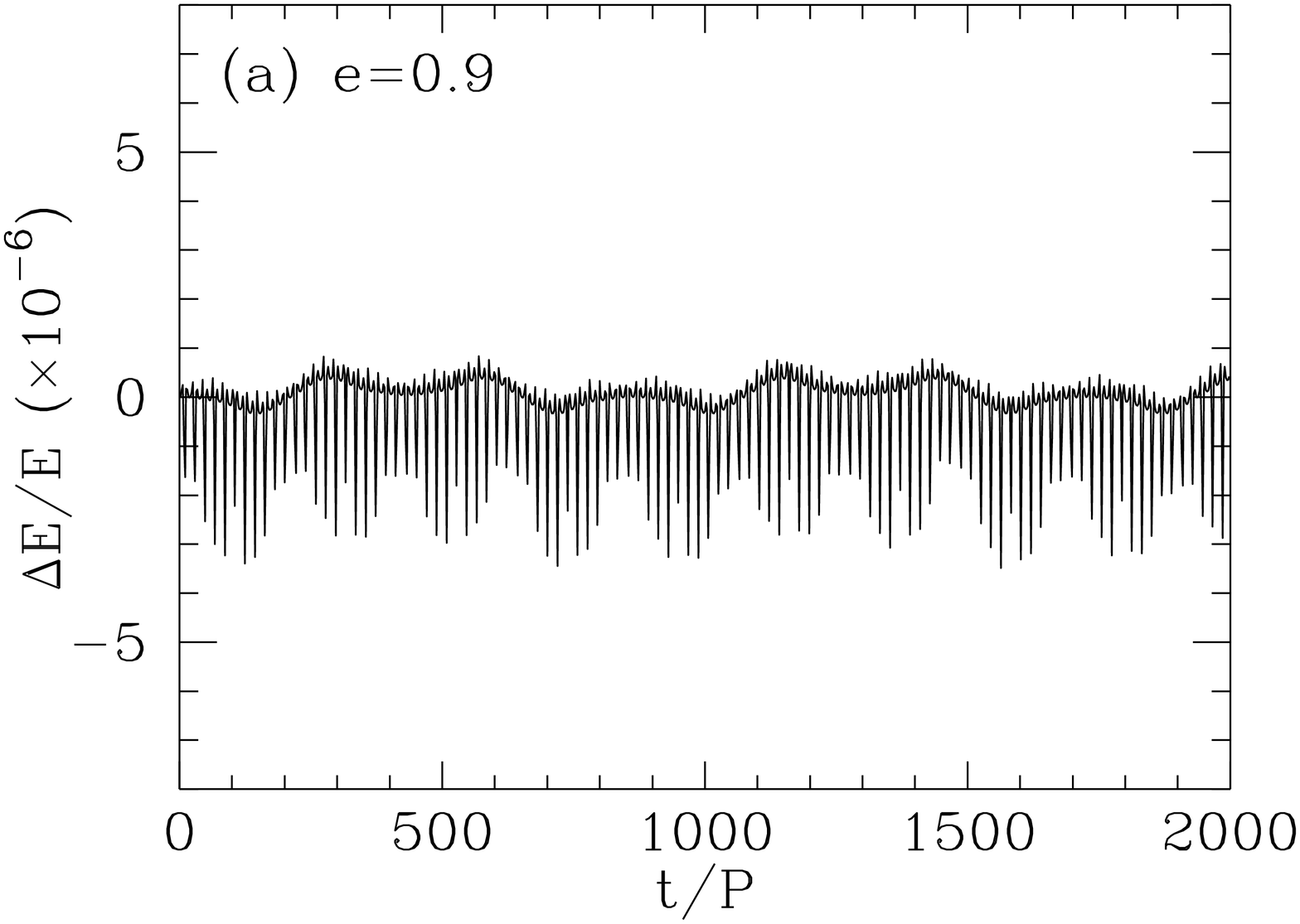}{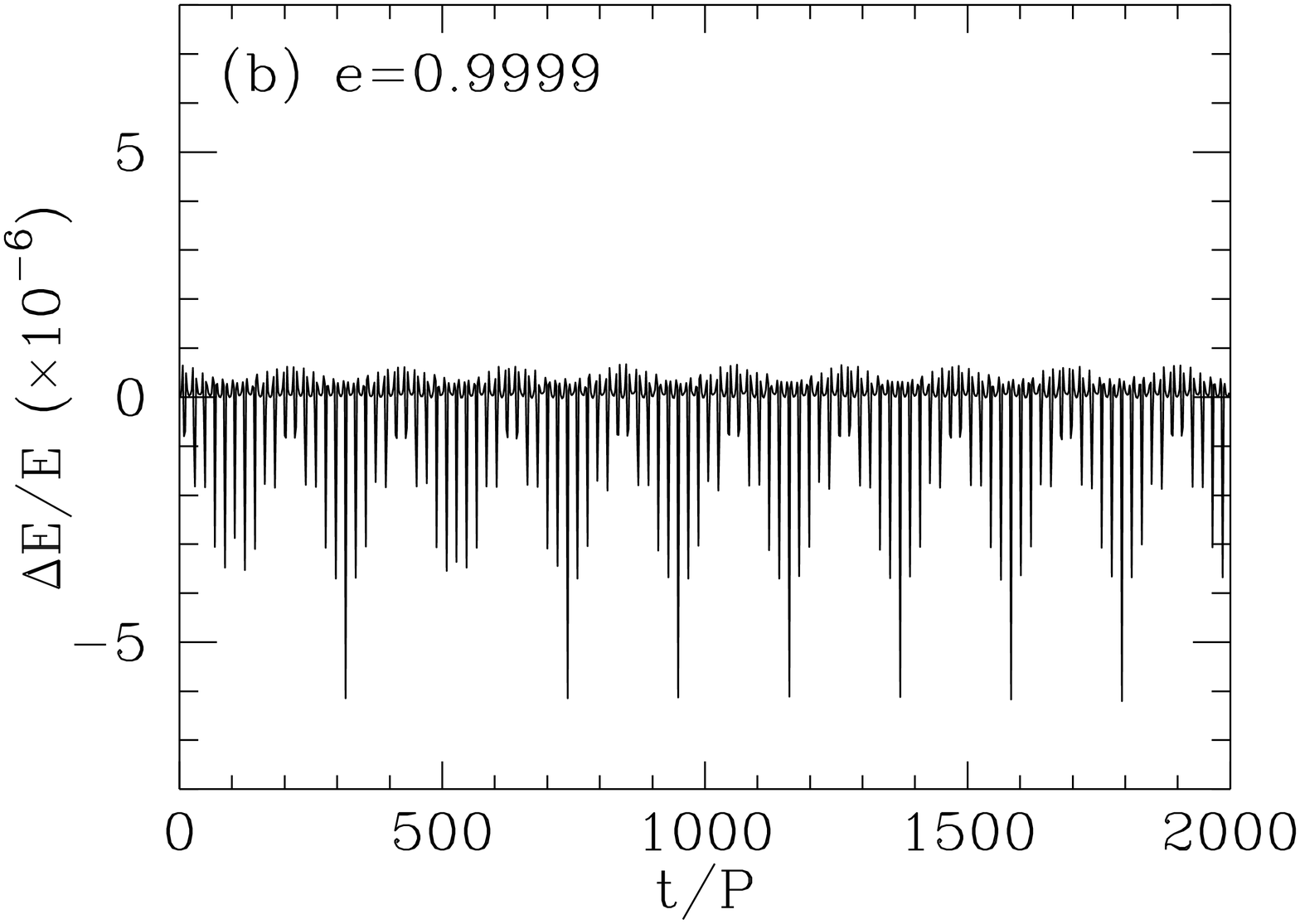}
\caption{Energy error $\Delta E/E$ of two integrations of the Kepler
problem using the multiple timescale symplectic integrator with $\vecF_{c,i}$.
The overall timestep $\tau_0 = P/2000$, $M_i = 2$,
$r_1 = \protect\sqrt{2}\,a$, and $r_i/r_{i+1} = \protect\sqrt{2}$.}
\end{figure}

One of the goals of this study is to develop a variable timestep integrator
for solar system integrations.
We have developed a second-order multiple timescale MVS integrator based on
the algorithm described in this section (see Levison \& Duncan 1994 for
another approach).
We are currently testing this integrator in detail.
Initial results indicate that the integrator is fast and accurate and has
all the desirable properties of the constant timestep symplectic integrators.

\end{document}